\documentclass[prl]{revtex4}

\usepackage{graphicx,t1enc}
\usepackage{epsfig}
\usepackage{amssymb,amsmath}

\begin{document}

\title{Inertial movements of the iris as the origin of post-saccadic oscillations}

\author{S. Bouzat$^{1*}$, M. L. Freije$^2$, A. L. Frapiccini$^2$,  G. Gasaneo$^2$}
\affiliation{$^1$Consejo Nacional de Investigaciones Cient\'{\i}ficas y T\'ecnicas, \\
Centro At\'omico Bariloche (CNEA), (8400) Bariloche, R\'{\i}o Negro, Argentina.
\\
$^2$Neufisur - Departamento de F\'{i}sica, Universidad Nacional del Sur - IFISUR, Bah\'{i}a Blanca (8000), Buenos Aires, Argentina.}

\thanks{email: bouzat@cab.cnea.gov.ar.}

\begin{abstract}
Recent studies on the human eye indicate that the pupil moves inside the eyeball due to deformations of the iris. Here we show that this phenomenon can be originated by inertial forces undergone by the iris during the rotation of the eyeball. Moreover, these forces affect the iris in such a way that the pupil behaves effectively as a massive particle. To show this, we develop a model based on the Newton Equation on the non-inertial reference frame of the eyeball. The model allows us to reproduce and interpret several important findings of recent eye-tracking experiments on saccadic movements. In particular, we get correct results for the dependence of the amplitude and period of the post-saccadic oscillations on the saccade size, and also for the peak velocity. The model developed may serve as a tool for characterizing eye properties of individuals.      
  
\end{abstract}

\maketitle

It has recently been shown that, during eyeball rotations, the iris suffers deformations that
can produce an effective motion of the pupil inside the eyeball \cite{holm2013,kim,holm,holm2016}.
This phenomenon is observed particularly at the end of the saccades \cite{sacdef,saccade}. Many high-quality videos (both professional and homemade) have been spread on the Internet in the last years showing details about this surprising issue \cite{videos}. The effect is usually ascribed to the interplay between the viscoelastic properties of the iris and the rotation of the eyeball, since this seems rather intuitive from the observations of the videos. However, a formal quantitative description is still lacking. 

While eye-tracking techniques are turning into an important tool for neuroscience, industry and marketing 
\cite{Duch}, recent experiments indicate that the motion of the pupil inside the eyeball can affect the measurements 
\cite{holm2016}. This is because such motion can be related to the observance of the so-called post-saccadic oscillations (PSO) \cite{holm2013}. Hence, the development of models analyzing pupil motion is important 
not only from the point of view of basic research, concerning the characterization of the eye physiology, but also for the interpretation of eye-tracking experiments. Concretely, models would help to shed light on the problem of distinguishing  information related to neural commands for eye motion, from data reflecting mechanical phenomena inside the eyeball.

With these considerations in mind, in this work we develop a model for the separate dynamics of eyeball and pupil during saccadic motion. The model helps us to answer important questions such as to what extent the motion of the pupil inside the eyeball can be related to the PSOs. Moreover, it allows us to understand the dependence of the PSO profiles on the saccade size reported in \cite{holm}, and how this is connected to the dependence of the peak velocity found in \cite{Otero:2008,MartinezC:2013}. Previously, mathematical models have been found useful for studying other aspects of saccadic motion \cite{Enderle:2010,Artal:2014,gasaneo2017,kliegl2},
microsaccades \cite{chino1,chino2} and fixation \cite{liang,Merge,kliegl1}, but the questions posed here were not analyzed. 

{\em Model for the eyeball and iris motion during saccades.-} We propose a 
one-dimensional model in which the eyeball motion is described by a dynamical variable $x$ representing the angular position of the center of the cornea along a saccade. Meanwhile, a second variable $y$ represents the relative angular position of the  pupil center measured from $x$.

Assuming that the eyeball is driven by the extra ocular muscles in an overdamped way, we consider the dynamical equation
\begin{equation}
\label{xpunto}
\nu \, \dot{x}=F(t).
\end{equation} 
Here, $\nu$ is the viscosity acting on the eyeball and $F(t)$ is the force representing the action of the muscles. For simplicity we fix $\nu=1$ so that the force is scaled. Given that we are only interested in describing single saccades, we consider the initial condition $x(0)=0$ with no loss of generality. The anisotropies that may affect the motion on different directions can be modeled
by varying the characteristics of $F(t)$.

In order to describe the relative motion of the pupil, we assume that the iris is elastically linked to the eyeball and that its internal border (which defines the pupil) can oscillate driven by inertial forces induced by the motion of the eyeball. The idea is that such inertial forces could act directly on the inner part of the iris (or on other internal pieces of the eye linked to the iris) in such a way that the pupil centre behaves effectively as a massive particle. Hence, we consider the equation:
\begin{eqnarray}
\label{ypunto}
\ddot{y}+\gamma\,\dot{y}+k\, y&=&-\ddot{x}.
\end{eqnarray}
Here, $k$ is the effective elastic constant that tends to bring the center of the pupil to its rest position on the eyeball, while $\gamma$ measures an effective viscosity affecting the relative motion. Finally, $-\ddot{x}$ stands for the inertial force {\em felt} by the inner part of the iris on the reference frame of the eyeball. Note that the mass is set equal to $1$ with no loss of generality. The units of $k$ and $\gamma$ are chosen to express $x$ and $y$ in degrees and the time in milliseconds. As initial conditions, we consider $y(0)=\dot{y}(0)=0$ so that the iris is at rest at its relaxation position. 

Although the motion of the pupil may obey complex phenomena associated with three-dimensional movements and deformations, by fitting the parameters $k$ and $\gamma$ and the function $F(t)$, our approach aims at capturing as much as possible the effective one-dimensional dynamics registered by eye trackers. In most of our work we consider $k$ and $\gamma$ as constant parameters. This is enough to understand several aspects of the phenomenology of the PSO. However, at the end of the work we show that the consideration of $k$ and $\gamma$ as functions of $F(t)$ can lead to better descriptions. Other versions of damped oscillators have been used in \cite{Artal:2014} and \cite{gasaneo2017} to describe saccadic motion. However, no attempt was made to analyze the inertial effects on the relative motion of the iris. 

We consider the forcing profile as given by
\begin{equation}
F(t)=A\, t \, \exp\left[-\frac{t^\mu}{\tau^\mu}\right],      \label{forcing}
\end{equation}
where $A$, $\tau$ and $\mu$ are positive parameters. Note that $\tau$ is a time constant, while $A$ gives a global measure of the strength of the forcing. The particular functional form for $F(t)$ given in Eq. (\ref{forcing}) is chosen for two reasons. First, it has the suitable characteristics for describing the average muscle activity during a saccade: it starts from $F(0)=0$, then grows up to a maximal value and decays again to zero. More importantly, as we will show, by varying only the parameter $\tau$, $F(t)$ generates saccades whose maximum velocity grows with the saccade size $x_m$ as $x_m^\alpha$, with $\alpha\sim 1/2$, in close agreement with experimental results found in \cite{Otero:2008} (using EyeLink 2 Eye Tracker) and in \cite{gasaneo2017} (with EyeLink 1000). A preliminary analysis of data for accelerations and velocities suggests values of $\mu$ in the range of $2-3$. In this work we find enough to consider $\mu=2$, while a complete study of role of this parameter will be discussed elsewhere.

{\em Solutions for eyeball motion.-} By considering Eq. (\ref{forcing}) with $\mu=2$, we can integrate Eq. (\ref{xpunto}) to get $x(t)=\frac{1}{2} A \tau^2 \left(1-\exp[-\frac{t^2}{\tau^2}]\right)$. The saccade size $x_m$ is just $x_m\equiv\lim_{t\to\infty}x(t)=A \tau^2 /2$, while the maximal velocity of the eyeball yields ${\rm Max}[\dot{x}]=A\,\tau/\sqrt{2 e}$. Note that, at fixed $A$, the saccade size can be controlled by the parameter $\tau$. We have $\tau=\sqrt{2 x_m/ A}$. Thus, from now on we use $x_m$ instead of $\tau$ as a relevant parameter. Interestingly, the saccades generated by varying $x_m$ at fixed $A$ satisfy the relation ${\rm Max}[\dot{x}]=\sqrt{A\, x_m/e}$, as can be seen by equating $\tau$ from the formulas for $x_m$ and ${\rm Max}[\dot{x}]$. This is the relation between the maximal velocity and $x_m^{1/2}$ mentioned before. It should be noted, however, that the maximal velocity of the pupil registered in experiments would actually correspond to ${\rm Max}[\dot{x}+\dot{y}]$, not to ${\rm Max}[\dot{x}]$. This does not make a big difference concerning the power law involved, as we later show. 

{\em Results for pupil motion and PSO.-} Because of the particular form of $F(t)$, the analytical solution for Eq. (\ref{ypunto}) is not straightforward and, in general, may involve hypergeometric functions. This may be matter for further studies. Here, in order to analyze the phenomenology of the model, we solve Eq. (\ref{ypunto}) numerically with $\ddot{x}$ derived from the solution $x(t)$ given above. We focus on the analysis of families of saccades of different amplitudes performed by the same eye in a fixed direction, as those studied in the experiments in \cite{holm}. In order to generate a family of saccades with these characteristics, we consider our model with varying $x_m$ at constant $A, k$ and $\gamma$. As explained before, the condition of constant $A$ leads to the relation ${\rm Max}[\dot{x}]\sim \sqrt{x_m}$. Meanwhile, the consideration of constant $k$ and $\gamma$ indicates that the iris-eyeball interaction is the same for 
every $x_m$. 

\begin{figure}  
\includegraphics[width=0.9\textwidth]{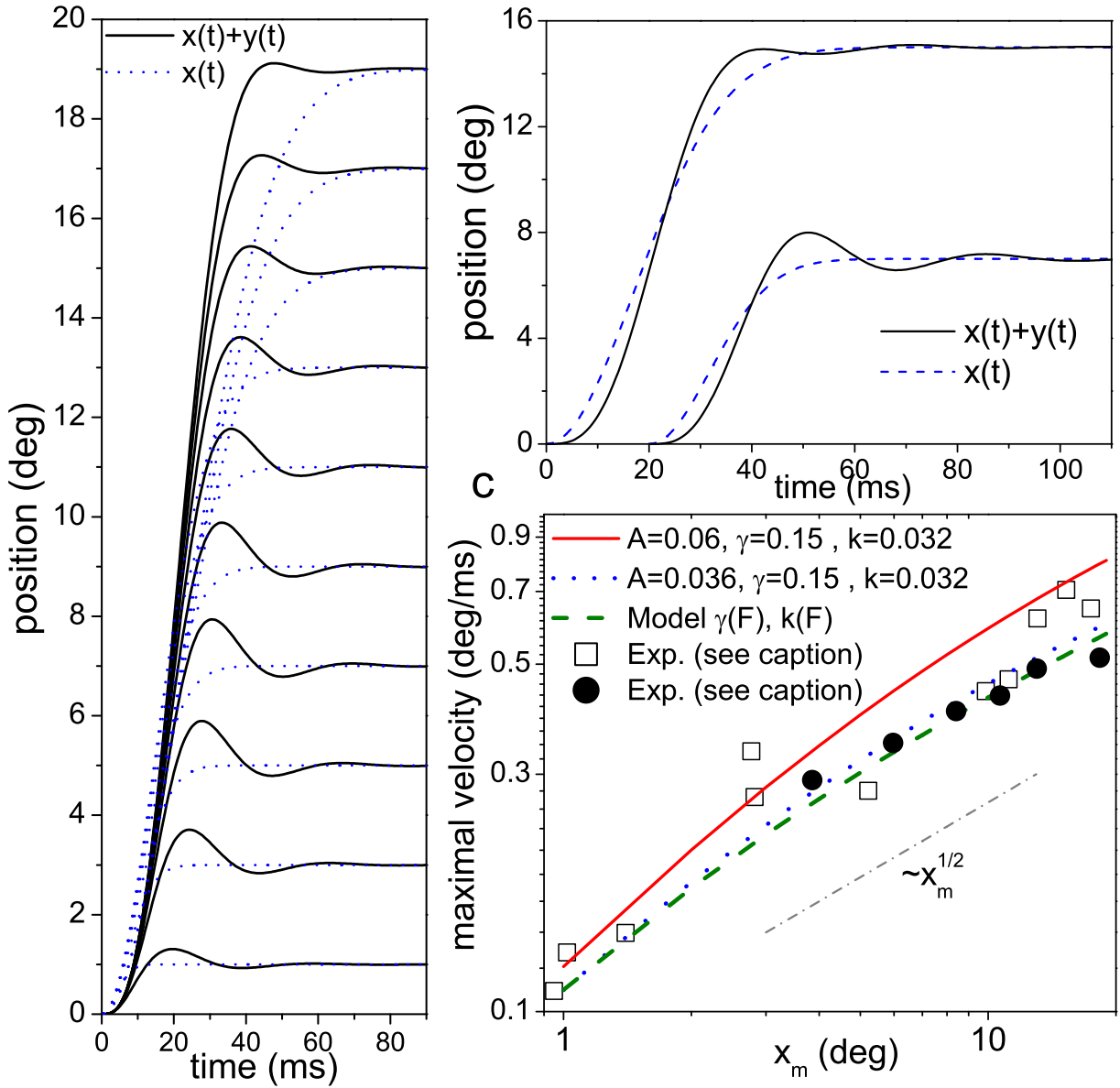}
\caption{(Color online) {\bf Pupil motion}. a) Family of saccades with constant $A=0.06$, $\gamma=0.15$, $k=0.032$ and varying $x_m$. b) Detail of eyeball ($x(t)$) and pupil ($x(t)+y(t)$) positions during saccades of sizes $x_m=7$ and $x_m=15$ calculated for $A=0.05$, $\gamma=0.1$ and $k=0.035$.  c) Maximal velocity vs. $x_m$. The red-solid and the blue-dotted lines correspond to families of saccades with constant $A, \gamma, k$ for the values indicated (the red solid line corresponds to the saccades in panel a). The white squares are results from (single) saccades from the same observer for a left eye moving to the left, taken from  experiments in \cite{gasaneo2017}. The black circles are our estimations using data recovered from the mean saccades shown in Fig.2 in \cite{holm} (pupil-corneal reflection signal) for the case of observer 3, left eye, abduction. The green dashed line corresponds to the model with force dependent parameters presented at the end of the work, with parameters as in Fig.\ref{figfit}.b. The dash-dotted segment indicates the $x_m^{1/2}$ behavior.}
\label{figgamcte}
\end{figure}

In Fig. \ref{figgamcte}.a we show a family of saccades of this type, while in Fig. \ref{figgamcte}.b we depict two saccades for another parameter set in order to show the details of the eyeball and pupil trajectories. The profiles found for $x(t)+y(t)$ are compatible with observations for pupil motion performed with EyeLink \cite{holm,gasaneo2017} and SMI \cite{holm,holm2016} eye trackers, exhibiting realistic shapes of PSO. Moreover, in agreement with what some results in \cite{holm2016} suggest, the model indicates that the pupil starts to move later than the eyeball, and it begins to oscillate before the eyeball reaches its stationary position. Even more, our results agree with the recent finding concerning the fact that the peak velocity of the pupil is larger than that of the corneal reflection \cite{holm2016}. All of these effects are easy to understand as caused by the interplay between the inertial and the elastic forces. Interestingly, our model also predicts that the amplitude of the PSO decreases with the saccade size at large $x_m$, as found in \cite{holm}. This will be discussed below. In Fig. \ref{figgamcte}.c  we show the dependence of the maximal velocity of the pupil ${\rm Max}[\dot{x}(t)+\dot{y}(t)]$ as a function of $x_m$ for two families of saccades with fixed $\gamma, k$ and $A$, together with data from experiments and from the model with force-dependent parameters described later. The results obtained with constant parameters are in good agreement with the experimental data, which have considerable dispersion for different saccades, directions and observers, as shown in Ref. \cite{gasaneo2017}. For $x_m\gtrsim 2$ deg, the calculated curves exhibit an approximately power-law behavior with exponent $\sim 1/2$.

\begin{figure}  
\includegraphics[width=0.9\textwidth]{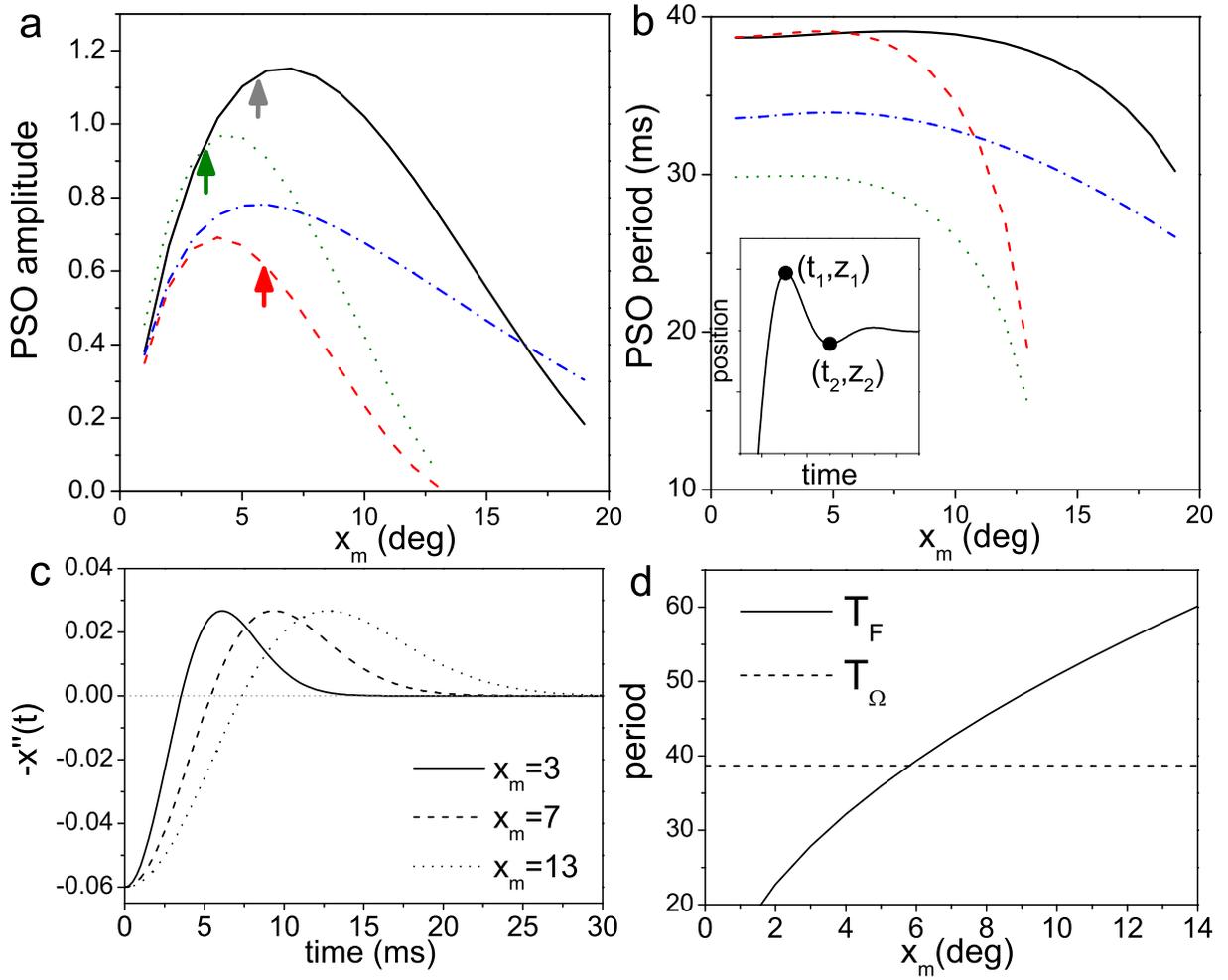}
\caption{(Color online). {\bf Amplitude and period of the PSO.} a) PSO amplitude as a function of $x_m$. The black solid line is for $A=0.06$, $\gamma=0.15$, $k=0.032$ (set in Fig.\ref{figgamcte}.a), the red dashed line is for $A=0.036$, $\gamma=0.15$, $k=0.032$, and the green dotted line for $A=0.06$, $\gamma=0.15$, $k=0.05$. The blue dash-dotted line is for the model with force dependent parameters with the values used in Fig.\ref{figfit}.b. The arrows indicate the value of $x_m$ for which $T_F$ and $T_\Omega$ match. b) PSO period as a function of $x_m$ for the same calculations as in panel (a). The inset sketches the method for calculation of the PSO amplitude and period. c) Inertial force $-\ddot{x}(t)$ for three values of $x_m$ from the set in black solid line in panel (a). d) $T_f$ and $T_\Omega$ as functions of $x_m$ for the set in panel (c). 
\label{figPSO}}
\end{figure}

The studies in \cite{holm} show that the amplitude of the PSO decreases with the saccade size for $x_m\gtrsim 8\, {\rm deg}$. Moreover, for most of the subjects analyzed, the PSO amplitude exhibits a maximum as a function of the saccade size at a value of $x_m$ in the range $5-8$ deg. Thus, the amplitude grows with $x_m$ at small $x_m$. The experiments in \cite{holm} also indicate that the PSO period decreases monotonically with $x_m$ for $x_m\gtrsim 4$ deg. Although these results were obtained with pupil-minus-corneal reflection (p-CR) signals \cite{supmat}, according to our developments in Ref. \cite{supmat}, we expect similar behaviors for pupil signals. As we will show, our model reproduces all the mentioned results. To define the PSO amplitude and period we use the following procedure (see inset in Fig. \ref{figPSO}.b). We label as $(t_1,z_1)$ the time-space position of the first local maximum of the saccade profile, and as $(t_2,z_2)$ that of the first minimum. Then, we define the PSO amplitude as $z_1-z_2$ and the period as $2(t_2-t_1)$ \cite{notePSO}. Fig. \ref{figPSO}.a shows the amplitude of the PSO as a function of $x_m$ calculated for different parameter sets, and also for the model with force-dependent parameters. In all the cases we find a maximum of the PSO amplitude for a value of $x_m$ in the range $4-8$ deg. This is compatible with the experimental curves shown in \cite{holm} in Fig. 5 (for horizontal saccades) and Fig. 9 (for vertical saccades). In Fig.  \ref{figPSO}.b we show the period of the PSO as a function of $x_m$ for the same set of calculations analyzed in Fig. \ref{figPSO}.a. We see that for $x_m\lesssim 8$ deg the period is almost independent of $x_m$ and then it exhibits a clear monotonic decreasing. This is also compatible with the experimental results shown in Fig. 6 in \cite{holm}. 

The decreasing of the PSO amplitude with $x_m$ at large $x_m$ was interpreted in \cite{holm} and references therein as due to a {\em gentle breaking} of the eyeball motion. As we here show, our model suggests that the existence of a maximum of the PSO amplitude as a function of $x_m$ can be interpreted as a resonant-like phenomenon related to the matching of a characteristic time of the eyeball forcing with the natural period of oscillation of the iris inside it. Fig \ref{figPSO}.c shows the inertial force $-\ddot{x}(t)$ plotted as a function of $t$ for three values of $x_m$. The forcing profile for $x_m=7$ is the one that produces the maximal amplitude of PSO. It is not easy to understand this result, since this forcing profile seems to have nothing special. Its duration is {\em intermediate} between those of the other two profiles plotted. One physically grounded explanation arises by noticing that the shape for the $-\ddot{x}(t)$ curve resembles a sinusoidal oscillation for which we can define a characteristic period. For this, we compute the cosine Fourier transform of $\Theta(t)\ddot{x}(t)$ (with $\Theta(t)$ the Heaviside function), and then the mean Fourier frequency and its associated period, referred to as $T_F$. On the other hand, we consider the natural frequency of oscillation of $y(t)$, namely $\Omega=\sqrt{k-(\gamma_0^2/4)}$, and its associated period $T_\Omega=2 \pi/\Omega$ which is independent of $x_m$. In Fig. \ref{figPSO}.d. we plot $T_\Omega$ and $T_F$ vs. $x_m$ for the same parameters considered in Fig. \ref{figPSO}.c. It can be seen that both periods match approximately at $x_m\simeq6$ deg, which is close to the value of $x_m$ that maximizes the PSO. The arrows in Fig. \ref{figPSO}.a indicate the values of $x_m$ for which $T_\Omega$ and $T_F$ match for the three parameter sets considered. In all the cases, it is close to the one that maximizes the PSO amplitude. This is reminiscent of a resonant phenomenon, although the maximization does not occur for a perfect matching. Note that this interpretation does not contradict the idea of a gentle breaking at large $x_m$, but helps to understand the growth of the PSO amplitude with $x_m$ at small $x_m$. 

\begin{figure}  
\includegraphics[width=0.9\textwidth]{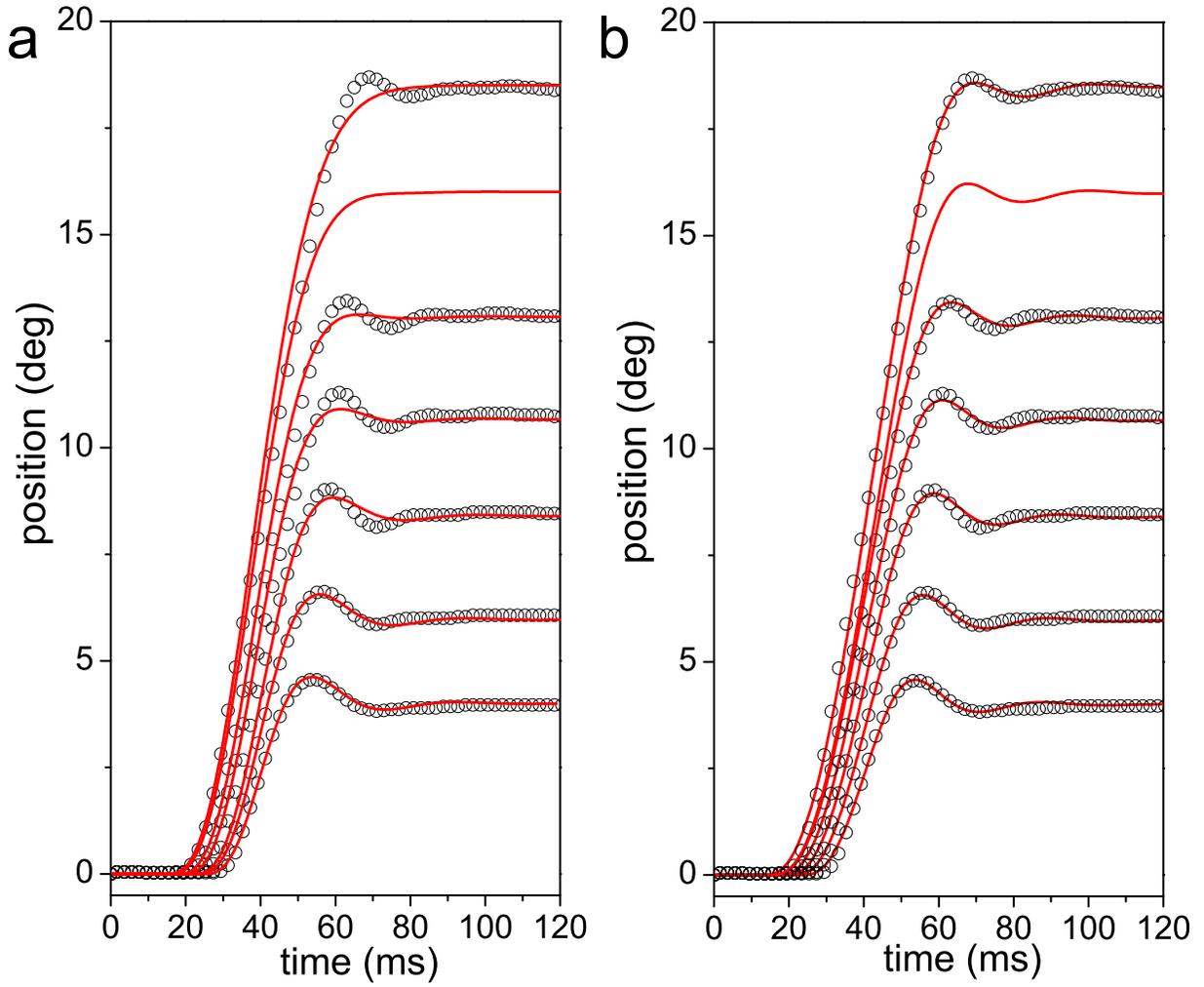}
\caption{(Color online). {\bf Fitting families of saccades from experiments}. In both panels, the open circles correspond to data recovered from experiments in \cite{holm} for the case indicated in Fig.\ref{figgamcte}.c. The solid red curves in (a) are our calculations for the model with constant parameters $A=0.04$, $k=0.032$, $\gamma=0.15$ for various values of $x_m$, while those in (b) are for the model with force-dependent parameters with $A=0.036$, $k_0=0.04$, $\gamma_0=0.14$, $c=0.5$, $d=3$. For the sake of completeness, in both panels we show calculations for $x_m=16$ although there are not experimental data for such saccade size.}
\label{figfit}
\end{figure}

{\em Fitting families of saccades: non constant parameters.-}
Although the model with constant parameters 
$A, k$ and $\gamma$ enables us to understand the dependence of the PSO amplitude and period on $x_m$, an accurate fitting of a family of experimental saccades may require additional sophistication. For instance, the consideration of a dependence of $k$ (or $\gamma$) on the relative position ($y(t)$) or on the relative velocity ($\dot{y}(t)$) or on the force ($F(t)$) may be in order. The development of a complete model would require a strong interplay between theory and experiments and an exhaustive analysis of many families of saccades with varying directions and luminance \cite{nyst2016A}, which is out of the scope of the present work. However, in order to give an example of a detailed fitting, we focus on reproducing the family of averaged horizontal saccades recovered from \cite{holm} that we show in Fig. \ref{figfit}. Importantly, as these saccades were obtained from p-CR signals, they may not describe accurately the actual motion of the pupil \cite{holm2016,supmat}, so that the fitted values of the parameters may not correspond exactly to those for the pupil dynamics. Nevertheless, the model can be used as a reasonable tool for fitting. Our results indicate that the model with constant $A, k$ and $\gamma$  overestimates the decay rate of the PSO amplitude with $x_m$, as shown in Fig  \ref{figfit}.a. Suitable PSO amplitudes at large saccade sizes can only be obtained by either overestimating the saccade velocity (by increasing $A$) or by overestimating the PSO amplitude for small $x_m$ (by decreasing $k$). In order to fit both the saccade velocities and the PSO profiles for all saccade sizes, we here consider the parameters $k$ and $\gamma$ as dependent on the force $F(t)$. This is meaningful from a physiological point of view since the forces exerted on the eyeball may not only rotate it but also produce smooth deformations which can change the interaction of iris with the crystalline and the other internal parts of the eye. We consider the particular forms $\gamma=\gamma[F(t)]=\gamma_0 \exp[-c F(t)]$ and $k=k[F(t)]=k_0 \exp[-d F(t)]$, with $c,d>0$. This represents a loosening of the eyeball-iris link with the force. In Fig. \ref{figfit}.b we show results for this model for a fixed set of parameters with varying $x_m$, together with the data recovered from \cite{holm}. The agreement is evident. It is worth remarking that the parameters $A, k, \gamma, c$ and $d$ are the same for all the saccades, and only $x_m$ changes. Still, we have not performed a fine tuning of the parameters, and so  the set considered is not expected to be the optimal one. The curve for the maximal velocity vs. $x_m$ for these parameters is shown in Fig. \ref{figgamcte}.c and agrees with the points calculated from the experimental saccades recovered. The results for the PSO amplitude and the period for the same calculations are shown in Figs. \ref{figPSO}.a and \ref{figPSO}.b, respectively. We see that decay of the PSO amplitude at large $x_m$ is slower than for the model with constant parameters, as we expected. 

{\em Final remarks.-} 
The simple model with constant parameters presented allows us to understand the PSO as a consequence of the inertial motion of the iris inside the eyeball. In particular, we get explanations for the dependence of the PSO profiles on the saccade size, and we show how this dependence is connected with that of the peak velocity. More elaborated versions of the model, such as the one with 
force-dependent parameters here considered, can provide accurate fittings of families of saccades. The model parameters could be fitted to determine individual characteristics of the eyes related to the viscoelastic link between iris and eyeball, and to the muscular force in different directions, with possible application to diagnostics \cite{aging}.

%Finally, it is interesting to compare the family of saccades shown in Fig. \ref{figgamcte}.a with the one
%for the model with constant parameters shown in Fig. \ref{figfit}.a, which differs only in the value of $A$. We see that a decreasing of $A$ leads to a reduction of the PSO amplitudes at large $x_m$, and also to a global decreasing of the saccade velocities. This is exactly what we observe in the results in \cite{holm} when comparing downward with upward saccades. This point, as well as the influence of the different model parameters will be further discussed in a separate publication. 

\section*{Acknowledgments}
The authors acknowledge support from Universidad Nacional del Sur (through PGI (24/F049)), CONICET (under Grant PIP 11220110100310) and from CNEA (all Argentinian institutions).

\begin{thebibliography}{99}
\bibitem{sacdef} A saccade is a fast binocular (coordinated) motion of the eyeballs that changes the fixation point. For more details, see, for instance, \cite{saccade} and \cite{Duch}.
\bibitem{notePSO} Our definitions of the PSO amplitude and period are different from those in \cite{holm}. However, this is not expected to affect the qualitative dependences on $x_m$. Our method is more easy to implement numerically since the localization of the second maximum considered in \cite{holm} is not always straightforward.
\end{thebibliography}
\end{document}